\documentclass[preprint,aps,amsmath,fleqn]{revtex4}

\usepackage{bm}
\usepackage[dvips]{graphicx}
\usepackage{color}
\usepackage{amsmath}
\usepackage{amssymb}

\newcommand{\eps}{\epsilon}
\newcommand{\kap}{\kappa}

\begin{document}

\title{Direct and indirect exciton mixture in double quantum wells}

\author{B. Laikhtman}
\affiliation{Racah Institute of Physics, Hebrew University, Jerusalem 91904, Israel}

\begin{abstract}
The exciton system in double quantum well is considered under condition when the ground state is the spatially indirect exciton. At high pumping growth of the exciton concentration can lead to so significant increase of the indirect exciton energy that becomes equal to the direct exciton energy. Then further increase of pumping leads to formation of mixed direct - indirect exciton phase. A rough estimate of the exciton energy in the mixed phase explains puzzling features of some recent exciton measurements. An experiment that would reveal main characteristic features of the mixed phase is suggested.
\end{abstract}

\maketitle

In the last two decades investigation of exciton system in double quantum wells motivated first by attempts to find Bose condensation found out many other quite interesting physical phenomena in this system.\cite{Butov02,Snoke02N,Snoke02S,Butov04a,Gorbunov06,Snoke11,High12a,Shilo13,Schinner13a,Alloing14,
Combescot17} The system of spatially indirect excitons has an extremely rich phase diagram. Depending on temperature and exciton density it can be in gas, liquid, correlated and ordered phases.\cite{Lozovik96,Lozovik97,Butov98,Butov04a,
Butov04b,Astrakharchik07,Lozovik07,High12b,Cohen16,Laikhtman09,Schinner13b,Remeika15,Suris16} This paper points out one more possibility. The exciton system can be a mixture of spatially direct and indirect excitons. And this state explains puzzling features of phase transition observed recently by Stern et al.\cite{Stern14,Misra18}

Direct (DX) and indirect (IDX) excitons in double well structure differ by location of the particles: both electron and hole are in the same well or they are in different wells. Typically holes are localized in one of the well and their tunneling to the other is hindered by mass a few times heavier than that of electron. If the electron levels in the wells are $\eps_{e1}$ and $\eps_{e2}$ and holes are located in well 1 then the energies of single direct and indirect excitons are
\begin{subequations}
\begin{eqnarray}
&& \eps_{DX} = \eps_{e1} - \eps_{DXb} \ ,
\label{eq:i.1a} \\
&& \eps_{IDX} = \eps_{e2} - \eps_{IDXb} \ ,
\label{eq:i.1b}
\end{eqnarray}
\label{eq:i.1}
\end{subequations}
where the reference energy is at the hole level and $\eps_{DXb}$ and $\eps_{IDXb}$ are binding energies of DX and IDX respectively. Confinement of electrons and holes in different quantum wells reduces the binding energy and $\eps_{DXb}>\eps_{IDXb}$. If $\eps_{e1}=\eps_{e2}$ then $\eps_{DX}<\eps_{IDX}$ and pumping leads to accumulation of DXs. However IDX, compared to DX, have longer recombination time that is controlled by the width and height of the barrier between the wells. Therefore interest to generation of dense exciton gas or exciton liquid is better to be met with IDX. To pump IDX it is necessary to make $\eps_{DX}>\eps_{IDX}$ that is usually done by manipulation of well
widths and external electric field to affect relative position of the electron levels so that
\begin{equation}
\eps_{e1} - \eps_{e2} > \eps_{DXb} - \eps_{IDXb} \ .
\label{eq:i.2}
\end{equation}
Then IDX become energetically favorable and the exciton system contains practically only IDX.

IDX, contrary to DX, have dipole moment and repulse each other. As a result the IDX energy in a system of many excitons becomes
$\eps_{IDX}+\eps_{int}$ where $\eps_{int}$ is the average interaction energy depending on the exciton concentration $n$. The blue shift of the IDX luminescence line by $\eps_{int}$ was detected in many experiments.\cite{Butov94,Negoita00,Butov01,Stern08,Schinner11,Cohen16} When the concentration is not very large and IDX system can be considered as non-ideal gas the interaction energy can be estimated according to\cite{Laikhtman09,Suris16}
\begin{equation}
\eps_{int} = 2\pi\Gamma(4/3)nr_{0}^{2}T \ ,
\label{eq:i.3}
\end{equation}
where $r_{0}=(e^{2}d^{2}/\kap T)^{1/3}$ is the minimal average classical distance between excitons, $d$ is the separation between the middle of the electron and hole wells and $\kap$ is the dielectric constant.
Eq.(\ref{eq:i.3}) is valid if the concentration is so small that $nr_{0}^{2}\lesssim1$. However, at temperature $T<2$ K and $nr_{0}^{2}=1$ Eq.(\ref{eq:i.3}) gives $\eps_{int}$ of a fraction of meV while typical blue shift at this temperature region is several meV. For $d=18$ nm and $\kap=12$  the minimal distance between excitons $r_{0}=67$ nm. Then the concentration corresponding to $nr_{0}^{2}=1$ is $2.2\times10^{10}$ 1/cm$^{2}$. At higher concentration the system is not in gas but in liquid state. Assuming that short range order in liquid is triangular lattice the interaction energy is estimated according to\cite{Laikhtman09}
\begin{equation}
\eps_{int} \approx 8 \ \frac{e^{2}d^{2}}{\kap} \ n^{3/2} \ .
\label{eq:i.4}
\end{equation}
For concentration $n=10^{11}$ 1/cm$^{2}$ Eq.(\ref{eq:i.5}) gives $\eps_{int}=9.8$ meV. That is the blue shift of several meV corresponds to the liquid state of the IDX system.

For typical liquids condensation and evaporation is the first order phase transition. But in these liquids the interaction between molecules is attractive at large distances with a short range repulsive core. The interaction between IDXs is quite different and in typical experimental situation transition from gas to liquid state is not a phase transition but a crossover.

Except dipole - dipole repulsion there is Van der Waals attraction between excitons that leads to formation of biexcitons in single well. However, at separation between electron and hole wells larger than the exciton Bohr radius $d>a_{X}=\hbar^{2}\kap/2m_{eh}e^{2}=7.1$ nm (electron and in-plane heavy hole masses are $m_{e}=0.067$ and $m_{hh}=0.14$, $m_{eh}=m_{e}m_{hh}/(m_{e}+m_{hh})$ and dielectric constant $\kap=12$) dipole - dipole repulsion dominates the interaction energy and $\eps_{int}(n)>0$ in the whole region of concentrations.\cite{Lozovik96} In other words, the interaction is repulsive at all distances. The same conclusion follows from calculation of biexciton binding energy in double quantum wells. Biexcitons don't exist if $d>1.8a_{X}$ for any $m_{e}/m_{hh}$ and if $d>a_{X}$ for $m_{e}/m_{hh}>0.4$.\cite{Zimmermann07,Meyertholen08,Lee09} Note also that dipole - dipole interaction falls off with distance as $1/r^{3}$ while Van der Waals interaction falls off as $1/r^{6}$. So that at large distances between excitons dipole - dipole repulsion dominates at any ratio $d/a_{X}$.

In case of repulsive potential between particles that monotonically falls off with the distance $\eps_{int}$ grows with the density. The energy of a particle defined as $\eps=(\partial E/\partial N)_{T}$, where $E$ is the internal energy of the system and $N$ is the number of particles, in liquid phase is practically equal to $\eps_{int}$ and its temperature dependence is very weak. Therefore chemical potential connected with $\eps$ by thermodynamic relation
\begin{equation}
\zeta - T \ \frac{\partial\zeta}{\partial T} = \eps
\label{eq:i.5}
\end{equation}
monotonically grows with concentration. In gas phase the main contribution to the chemical potential comes from thermal motion and it also grows with concentration. Eqs.(\ref{eq:i.3}) and (\ref{eq:i.4}) give
\begin{subequations}
\begin{eqnarray}
&& \zeta_{gas} = T \ln\frac{\pi\hbar^{2}n}{2m_{X}T} + \pi\Gamma(1/3)Tnr_{0}^{2} \ ,
\label{eq:i.6A} \\
&& \zeta_{liquid} \approx 8 \ \frac{e^{2}d^{2}}{\kap} \ n^{3/2} \ .
\label{eq:i.6b}
\end{eqnarray}
\label{eq:i.6}
\end{subequations}
where $m_{X}=m_{e}+m_{hh}$ is the exciton mass. The conclusion is the following: at large enough ratio $d/a_{X}$ that usually takes place in experiment the interaction between IDXs is repulsive and monotonically falls off with the distance. In this case the chemical potential of the system is a monotonically growing function of the concentration and the main condition of coexistence of gas and liquid phases $\zeta_{gas}=\zeta_{liquid}$ cannot be met. That is transition from gas to liquid state with growth of IDX concentration is not a phase transition but a crossover.

Basically, the conclusion seems to be rather simple. For phase separation an attraction between of the particles in one of the phases is necessary. If interaction between particles is dominated by repulsion at all concentrations the phases are mixed and cannot be distinguished.

In the presence of dipole - dipole repulsion between IDX condition of stability of IDX system (\ref{eq:i.2}) has to be modified and it becomes
\begin{equation}
\eps_{int} < \eps_{DX} - \eps_{IDX} = (\eps_{e1} - \eps_{e2}) - (\eps_{DXb} - \eps_{IDXb}) \ .
\label{eq:i.7}
\end{equation}
If condition (\ref{eq:i.7}) is violated DX states appear more energetically favorable compared to IDX and further increase of pumping leads to creation of DXs keeping IDX concentration constant. That is the system appears in mixed DX - IDX phase.

\begin{figure}[h]
\includegraphics[scale=0.8]{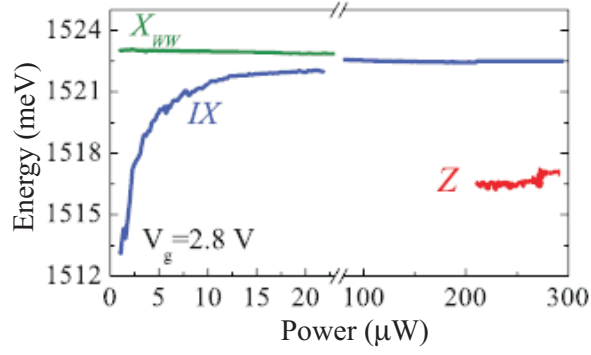}
\caption{\label{fig:lines}Luminescence peak positions depending on laser power in a GaAs/AlGaAs structure with $d=18$ nm at $T=1.5$ K.\cite{Stern14} $X_{WW}$ is the DX line, $IX$ is the IDX line. The two lines at power higher than $100\mu$W are luminescence of pure IDX phase and mixed DX-IDX phase ($Z$ line).}
\end{figure}

This behavior of the exciton system can be compared with evolution of the position of luminescence lines with growth of pumping observed by Stern at al.\cite{Stern14}, Fig.\ref{fig:lines}. At pumping right below the phase transition the blue shift
reaches $\approx$12 meV. Above estimates show that even at concentration below the phase transition the system is already in liquid state. A remarkable feature of the of the phase transition is that the new phase comes about when the IDX line approaches DX line. There is no increase of the blue shift with increase of the pumping above the transition point that indicates ceasing of further increase of IDX concentration. These features suggest that the phase transition is actually formation of mixed DX - IDX phase. The luminescence line at high pumping power (not $Z$ line) can be understood as luminescence of blue shifted IDX.

If the IDX concentration is $1.1\times10^{11}$ cm$^{-2}$ the average distance between them 29 nm is four times larger than the exciton radius $a_{X}=7.1$ nm. That is DX concentration can significantly surpass IDX concentration before their wave function start to overlap leading to a strong short range repulsion.

Direct excitons don't have dipole moment that is crucial to the structure of the mixed phase. In this phase the Van der Waals attraction between DXs and between DXs and IDXs dominates leading to formation of biexcitons and larger complexes.\cite{Usukura99}

\begin{figure}[h]
\includegraphics[scale=0.6]{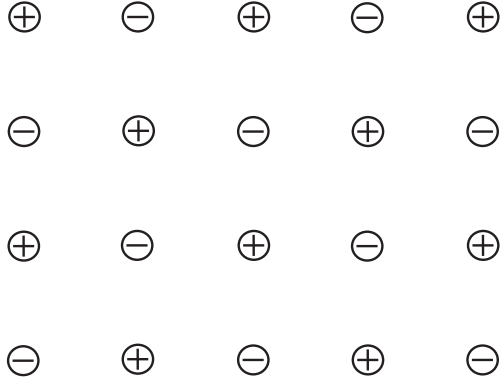}
\caption{\label{fig:e-h_lattice}Multi-exciton complex forms electron - hole lattice. Dashed lines separate unit cell.}
\end{figure}

Biexcitons in quantum well have a structure close to a square with electron wave function concentrated at the ends of one of the diagonals and hole wave function concentrated at the ends of the other one.\cite{Singh96} Although the exact square structure underestimate smearing of the electron and hole wave functions (zero point fluctuations)\cite{Denschlag99,Riva02} it correctly reflects the symmetry of the system where it is impossible to distinguish electron - hole pairs belonging to different excitons. Similarly, if the DX concentration is larger than IDX concentration DXs liquid has a short range order of many such squares, i.e., square electron - hole lattice, Fig.\ref{fig:e-h_lattice}. Sites of the lattice are positions of the maxima of electron and hole wave functions that are significantly smeared around. An IDX in such cluster of DXs can be considered as a defect and its structure is different
from the structure of free IDX by nonzero average lateral distance between electron and hole. This distance results from admixture of high angular momentum states to s-wave function of free IDX that is coming from Van der Waals interaction at large distances and even stronger at short distances between excitons. Attraction between DX  and DX - IDX leads to separation of the mixed DX - IDX phase from pure IDX phase.

The energy of the luminescence line of DX - IDX cluster can be estimated based on the binding energy of biexcitons in single quantum well. The binding energy of biexciton comes from attraction of one electron - hole pair by the other pair. In experiment this binding energy shows up as a red shift of the biexciton luminescence line compared to DX line. Different numerical methods were used for calculation of biexciton binding energy in single quantum well.\cite{Usukura99,Singh96,Denschlag99,Riva02,Kleinman83,Liu98} However, theoretical values are noticeably smaller than the latest measured value that is 1.5 - 2.1 meV depending on the well width.\cite{Birkedal96} Therefore the further estimate is based on the experimental value. A recombining electron - hole pair in a cluster Fig.\ref{fig:e-h_lattice} is attracted by six nearest neighbors compared to two neighbors in biexciton. So it is possible to expect that the red shift of the cluster luminescence line is about three times larger than that of biexciton. This result is close to the red shift of $Z$ line in Fig.\ref{fig:lines} compared to DX or blue shifted IDX line. This picture explains also spatial separation of the sources of $Z$ line and the other line.\cite{Stern14,Misra18}

Shift of $Z$ line linearly with the gate voltage shows that this line is due to recombination of IDX.\cite{Stern14} At low temperature this recombination requires tunneling of the electron to the hole well.\cite{Shilo13,Mazuz-Harpaz17} With growth of temperature transition of thermally activated electrons across the barrier becomes more prominent and reduces the IDX life time. As a result the critical IDX concentration can be maintained only with increase of pumping power. This is consistent with general shape of the line separating two phases in the phase diagram in the power - temperature plane.\cite{Misra18}

Manipulation with the gate voltage suggests also other experiments that directly confirm the nature of the new phase. With increase of the gate voltage the energy separation between electron states in the two wells, $\eps_{e1}-\eps_{e2}$, grows. Then critical concentration necessary for equality of the DX and IDX energies,
\begin{equation}
\eps_{IDX} + \eps_{int} = \eps_{DX} \ ,
\label{eq:i.6}
\end{equation}
and formation of the mixed phase becomes larger. If the exciton concentration is proportional to the pumping power $P$ and $\eps_{e1}-\eps_{e2}$ is proportional to the gate voltage $V_{g}$ then the critical pumping power $P_{c}$ has to grow as $V_{g}^{2/3}$. Strictly speaking, condition for formation of the mixed phase can include also exciton cluster binding energy that can weakly depend on the cluster size, i.e., on the DX concentration. At very high gate voltage the IDX concentration necessary for Eq.(\ref{eq:i.6}) becomes so large that the average distance between IDXs becomes comparable or even smaller than the exciton radius $a_{X}$. In this case increase of the pumping power leads rather to destruction of IDXs than to the formation of the mixed phase and the luminescence line dependence on power has to be qualitatively different from Fig.\ref{fig:lines} at high power.

In conclusion, in system of indirect excitons in double quantum well high pumping leads to increase of the exciton concentration and growth of the exciton energy due to dipole - dipole repulsion of the excitons. With increase of the concentration a crossover of the gas phase of excitons to liquid phase takes place. The concentration can reach the threshold when the indirect exciton energy becomes equal to the energy of direct excitons. Then further increase of pumping leads to growth of the  direct exciton concentration leaving the concentration of indirect excitons constant. Van der Waals attraction between direct excitons and between direct and indirect excitons leads to formation of a mixed phase of direct and indirect excitons. Luminescence of the mixed phase explains a puzzling $Z$ line in recent experiments. Another experiment is suggested that would explicitly confirm formation of the mixed phase.

I am thankful to M. Stern for a comment.

\end{document}